\documentclass[11pt]{article}
\usepackage{amsmath,amssymb,color,graphics,epsfig,cite,amsmath}

\usepackage{amsfonts}

\newcommand{\be}{\begin{equation}}
\newcommand{\ee}{\end{equation}}
\newcommand{\bea}{\setlength\arraycolsep{2pt} \begin{eqnarray}}
\newcommand{\eea}{\end{eqnarray}}

\def\0{{\sst{(0)}}}
\def\1{{\sst{(1)}}}
\def\2{{\sst{(2)}}}
\def\3{{\sst{(3)}}}
\def\4{{\sst{(4)}}}
\def\5{{\sst{(5)}}}
\def\6{{\sst{(6)}}}
\def\7{{\sst{(7)}}}
\def\8{{\sst{(8)}}}
\def\sst#1{{\scriptscriptstyle #1}}
\def\oneone{\rlap 1\mkern4mu{\rm l}}

\begin{document}

\begin{center}
{\Large {\bf Four-charge Rotating AdS Black Holes\\ in Gauged STU Supergravity}}

\vspace{20pt}

Liang Ma and H. L\"u

\vspace{10pt}

{\it Center for Joint Quantum Studies, Department of Physics,\\
School of Science, Tianjin University, Tianjin 300350, China }

\vspace{40pt}

\underline{ABSTRACT}
\end{center}

We present an exact solutions for non-extremal rotating AdS black holes with four independent charges in four-dimensional gauged STU supergravity. The scalar potential introduced by gauging breaks the hidden symmetry underlying solution-generating methods in the ungauged theory, making the construction of such solutions a longstanding challenge. We analyse the global properties of the solution and derive its thermodynamic quantities. We further take the BPS limit and show that the resulting entropy agrees with the known microscopic result obtained from the dual conformal field theory.

\vfill {\footnotesize maliang0@tju.edu.cn\ \ \ mrhonglu@gmail.com}


\thispagestyle{empty}
\pagebreak


\newpage

\section{Introduction}

The AdS/CFT correspondence \cite{Maldacena:1997re} elevates the importance of gauged supergravities in the top-down approach to the gauge-gravity duality. Gauged STU supergravities in five and four dimensions are particularly important in this context, as they are consistent and considerably simpler supersymmetric truncations of maximal gauged supergravities, retaining the Cartan subgroup of the corresponding gauge symmetry and thereby capturing the Abelian R-charge sector of the dual conformal field theory (CFT). The charges carried by the black holes are therefore identified with the R-charges of the corresponding dual CFT. However, electrically charged static anti-de Sitter (AdS) black holes \cite{Behrndt:1998jd, Duff:1999gh, Sabra:1999ux} generally develop naked singularities in the BPS limit. By contrast, regular magnetically charged static BPS AdS black holes do exist
\cite{Cacciatori:2009iz,Benini:2015eyy}. Two approaches have been explored to address this issue. One is the bubbling AdS construction, which replaces singular configurations with smooth horizonless BPS geometries \cite{Lin:2004nb,Chong:2004ce,Chen:2007du}. The other is to introduce rotation. Charged BPS rotating AdS black holes in gauged STU supergravities have also been constructed using the G-structure approach \cite{Gutowski:2004yv,Cacciatori:2008ek, Hristov:2019mqp}. Nevertheless, BPS solutions have vanishing temperature and therefore do not capture the full thermodynamic behaviour of finite-temperature black holes. Moreover, in theories with higher-derivative and Chern-Simons interactions, the evaluation of black hole entropy may require the extension of the standard Wald formalism, making a consistent treatment of the zero-temperature limit particularly subtle \cite{Tachikawa:2006sz,Ma:2022nwq}.

The construction of charged rotating AdS black holes in five-dimensional gauged supergravity followed a relatively systematic progression, from the ungauged solution \cite{Cvetic:1996xz} to gauged AdS black holes \cite{Cvetic:2004hs, Chong:2005hr} and eventually to the general three-charge solution \cite{Wu:2011gq}. By contrast, progress in four dimensions has been more gradual, and
a general rotating AdS black hole with four independent charges has remained elusive,\footnote{The solution was first reported by Shuang-Qing Wu in a gravity conference (June, 2015) in Hangzhou, China, but still unpublished, and we are unaware of any talk notes or slides available in the public.}  despite the construction of several special cases over years \cite{Chong:2004na,Chow:2010sf,Chow:2010fw, Wu:2011zzh,Chow:2013gba}. In this paper, we present such an exact solution for rotating non-extremal AdS black holes with four independent charges, and investigate its global properties, the black hole thermodynamics, and the BPS limit.

\section{Four-charge rotating AdS black holes}

\subsection{The theory}

Four-dimensional ungauged STU supergravity is a consistent truncation of maximal ${\cal N}=8$ supergravity, Its bosonic field consists of the metric, four Abelian gauge fields, and three complex scalars $z_i=\chi_i + {\rm i} e^{-\varphi_i}$, with ($i=1,2,3$.) The model possesses a global symmetry of $SL(2,\mathbb R)^3$, corresponding to the strong/weak coupling (S-)duality, target spacetime (T-)duality and U-duality \cite{Duff:1995sm}. Upon stationary (timelike) Kaluza-Klein (KK) reduction \cite{Cremmer:1998em} to three dimensions, the theory can be reformulated as a nonlinear sigma model with the target space given by the coset $SO(4,4)/SL(2,\mathbb R)^4$ \cite{Chow:2014cca}. This hidden symmetry can be exploited to construct exact black hole solutions: by taking the Kerr black hole as a seed and applying suitable solution-generating transformations, one can obtain the general four-charge non-extremal rotating black hole \cite{Cvetic:1996kv,Chong:2004na}. The STU also admits a gauged version. At the level of the bosonic Lagrangian, gauging introduces a scalar potential, which gives rise to an AdS vacuum with cosmological constant $\Lambda=-3g^2$, where $g>0$ denotes the gauge coupling. Adopting the convention of \cite{Chong:2004na}, the Lagrangian takes the form
\bea
\mathcal{L}_4&=&R {*\oneone}-\frac{1}{2} {*d\varphi_i}\wedge d\varphi_i-\frac{1}{2}e^{2\varphi_i} {*d\chi_i}\wedge d\chi_i-\frac{1}{2}e^{-\varphi_1}\Big(
e^{\varphi_2-\varphi_3} {*F}_{\2,1}\wedge F_{\2,1}\cr
&&+e^{\varphi_2+\varphi_3} {*F}_{\2,2}\wedge F_{\2,2}+e^{-\varphi_2+\varphi_3} {*\mathcal{F}}_{\2}^1\wedge \mathcal{F}_{\2}^1
+e^{-\varphi_2-\varphi_3} {*\mathcal{F}}_{\2}^2\wedge \mathcal{F}_{\2}^2
\Big)\cr
&&-\chi_1\big(F_{\2,1}\wedge \mathcal{F}_{\2}^1+F_{\2,2}\wedge \mathcal{F}_{\2}^2\big)
+g^2\sum_{i=1}^3\big(2\cosh\varphi_i+\chi_i^2e^{\varphi_i}\big){*\oneone}\,,\cr
F_{\2,1}&=&dA_{\1,1}-\chi_2d\mathcal{A}_{\1}^2\,,\qquad
F_{\2,2}=dA_{\1,2}+\chi_2d\mathcal{A}_{\1}^1-\chi_3dA_{\1,1}+
\chi_2\chi_3d\mathcal{A}_{\1}^2\,,\cr
\mathcal{F}_{\2}^1&=&d\mathcal{A}_{\1}^1+\chi_3d\mathcal{A}_{\1}^2\,,\qquad
\mathcal{F}_{\2}^2 = d\mathcal{A}_{\1}^2\,.\label{lag}
\eea
The gauge coupling $g$ enters the Lagrangian only through the scalar potential. In the ungauged limit $g=0$, the above Lagrangian can be derived directly from six-dimensional string theory compactified on two-torus \cite{Duff:1995sm}, retaining both the two winding modes $(A_{\1,1}, A_{\1,2})$ and two KK modes $({\cal A}_{\1}^1,{\cal A}_{\1}^2)$. To establish a direct correspondence with the $U(1)^4$ gauged supergravity obtained from a consistent truncation of the $SO(8)$ maximum gauged supergravity \cite{Cvetic:1999xp,Azizi:2016noi}, the two Maxwell fields $(A_{\1,1}, {\cal A}_{\1}^1)$ must both be Hodge-dualized. Although electromagnetic duality is generally not a symmetry of the full gauged supergravity, the corresponding duality transformations can still be implemented for the $U(1)$ gauge fields in the purely bosonic sector, with the fermions set to zero. Consequently, black holes carrying four electric charges in $U(1)^4$ gauged supergravity admits an equivalent description in terms of two electric and two magnetic charges.

In the presence of a scalar potential, the solution-generating techniques based on the hidden symmetries of the ungauged theory are no longer directly applicable. Nevertheless, guided by the known neutral rotating AdS black hole \cite{Carter:1968ks} and four-charge rotating solutions \cite{Cvetic:1996kv,Chong:2004na} of the ungauged theory, several rotating AdS black holes with restricted charge configurations were constructed using suitably motivated ans\"atze. These included the pairwise-equal charges \cite{Chong:2004na, Chow:2013gba}, as well as single-charge and two-charge truncations \cite{Chow:2010sf,Chow:2010fw}. Independently, supersymmetry-based G-structure methods have yielded BPS rotating AdS black holes with four independent charges \cite{Cacciatori:2008ek,Hristov:2019mqp}.

\subsection{The solution}

Here, we present the general four-charge non-extremal rotating AdS black hole. Although the solution is rather involved, it can be expressed in a relatively compact form. We present the metric, the four $U(1)$ fields, and the scalar fields separately. To write the metric in a compact form, we introduce following shorthand notations:
\bea
&&s_i=\sinh\delta_i\,,\qquad c_i=\cosh\delta_i\,,\qquad r_i=r+2\mu s_i^2\,,\cr
&&c_{i_1\cdots i_n}=c_{i_1}\cdots c_{i_n}\,,\qquad s_{i_1\cdots i_n}=s_{i_1}\cdots s_{i_n} \,,\qquad r_{i_1\cdots i_n}=r_{i_1}\cdots r_{i_n}\,.
\eea
We define $\Xi=1-a^2g^2$, $u=a\cos\theta$ and $\Delta = r^2 - 2\mu r + a^2$ and
\bea
\Delta_r&=&\Xi\,\Delta+g^2 F\,,\qquad \Delta_\theta = 1-g^2u^2\,,\qquad N=F-a^2 \sin^2\theta \Delta\,,\qquad
P=r_{1234}\,,\cr
K&=&2r^2+2\mu r\sum_{i=1}^4s_i^2+8\mu^2c_{1234}s_{1234}-4\mu^2(s_{123}^2+s_{124}^2
+s_{134}^2+s_{234}^2+2s_{1234}^2)\,,\cr
L&=&2\mu\big[rc_{1234}-(r-2\mu)s_{1234}\big]\,,\quad
F=P+a^2K+a^4,\quad W^2=P+u^2K +u^4.\label{function 4}
\eea
The black hole metric can then be expressed in a simple form:
\be
ds^2 = -\frac{W\Delta_r \Delta_\theta}{\Xi N} dt^2 +
\frac{N\sin^2\theta}{\Xi W} \Big(d\phi-\frac{a \Delta_\theta\,L}{N} dt\Big)^2+W\Big(\frac{dr^2}{\Delta_r}+\frac{d\theta^2}{\Delta_\theta}\Big)\,.\label{metric}
\ee
The square root of the determinant of the metric is simply $W \sin\theta/\Xi$. We introduce $\bar{i}$ to denote the complementary triple of indices obtained by excluding $i$ from the set $\{1,2,3,4\}$. Explicitly. $\bar{1}=234$, $\bar{2}=134$, $\bar{3}=124$, and $\bar{4}=123$. We then define
\bea
\alpha_i&=&s_ic_{\bar{i}}-c_is_{\bar{i}}\,,\quad
e_i=c_{\bar{i}}s_{\bar{i}}(c_i^2+s_i^2)-c_is_i\Bigg(\sum_{\substack{j<k,\\ j,k\neq i}}s_j^2s_k^2+2s_{\bar{i}}^2\Bigg)\,,\quad \mathcal{R}_i=\prod_{j\neq i}r_j+ru^2\,,\cr
\mathcal{H}_i&=&c_is_i\Big\{
a^2(rr_i+u^2)+P+u^2\big[r^2+2\mu r\sum_{j\neq i}s_j^2-4\mu^2s_{\bar{i}}^2\big]
\Big\}\cr
&&+4\mu^2u^2c_{\bar{i}}s_{\bar{i}}s_i^2+2\mu  e_i(a^2r_i-ru^2)\,.
\eea
The four $U(1)$ gauge fields can be written in the following compact form
\bea
A_{\1,1}&=&\frac{2\mu\cos\theta}{\sqrt{\Xi}W^2}\Bigg\{
a\Big[\Delta_\theta\big((rr_1+u^2)\alpha_1+2\mu r_1c_1s_{\bar{1}}\big)+\alpha_1g^2W^2\Big]dt-\mathcal{H}_1d\phi
\Bigg\}\,,\cr
A_{\1,2}&=&\frac{2\mu}{\sqrt{\Xi}W^2}\Bigg\{\Delta_\theta(c_2s_2\mathcal{R}_2+2\mu u^2e_2)dt
-a\sin^2\theta(\alpha_2\mathcal{R}_2+2\mu u^2c_2s_{\bar{2}})d\phi
\Bigg\}\,,\cr
\mathcal{A}_{\1}^1&=&\frac{2\mu\cos\theta}{\sqrt{\Xi}W^2}\Bigg\{
a\Big[\Delta_\theta\big((rr_3+u^2)\alpha_3+2\mu r_3c_3s_{\bar{3}}\big)+\alpha_3g^2W^2\Big]dt-\mathcal{H}_3d\phi
\Bigg\}\,,\cr
\mathcal{A}_{\1}^2&=&\frac{2\mu}{\sqrt{\Xi}W^2}\Bigg\{\Delta_\theta(c_4s_4\mathcal{R}_4+2\mu u^2e_4)dt
-a\sin^2\theta(\alpha_4\mathcal{R}_4+2\mu u^2c_4s_{\bar{4}})d\phi
\Bigg\}\,.\label{U1 gauge field}
\eea
As discussed above, in the theory described by the Lagrangian \eqref{lag}, the two Maxwell fields $(A_{\1,1},\mathcal{A}_{\1}^1)$ carry magnetic charges, whereas $(A_{\1,2},\mathcal{A}_{\1}^2)$ carry electric charges. If we Hodge-dualize the former pair, all four gauge fields will carry electric charges, and their expressions can be written in the form of the last term, with the index 4 replaced by 1, 2 and 3 respectively. In this duality frame, the functions ${\cal H}_i$ are no longer needed. Alternatively, if we Hodge-dualize the latter pair, all four gauge fields can be expressed in the form of the first term, with the index 1 resplaced by 2, 3 and 4 respectively. In this case, the quantities $e_i$ are not necessary. Finally, the three dilatons and three axions are given by
\bea
e^{\varphi_1}&=&\frac{r_{13}+u^2}{W}\,,\qquad e^{\varphi_2}=\frac{r_{23}+u^2}{W}\,,\qquad e^{\varphi_3}=\frac{r_{12}+u^2}{W}\,,\cr
\chi_1&=&\frac{2\mu (c_{13}s_{24}-c_{24}s_{13}) u}{r_{13}+u^2}\,,\quad
\chi_2= \frac{2\mu (c_{14}s_{23}-c_{23}s_{14}) u}{r_{23}+u^2}\,,\quad
\chi_3=\frac{2\mu (c_{12}s_{34}-c_{34}s_{12}) u}{r_{12}+u^2}\,.
\eea

\subsection{Thermodynamics}

The solution is asymptotic to the non-rotating AdS spacetime expressed in spheroidal coordinates. Therefore, its thermodynamic quantities can be calculated straightforwardly using standard methods. For sufficiently large values of $\mu$, the solution admits an event horizon, which is located at $r=r_+$, where $r_+$ denotes the largest root of $\Delta_r=0$.  The entropy, temperature, and angular velocity of the black hole are given, respectively, by
\be
S=\frac{\pi}{\Xi}\sqrt{F(r_+)}\,,\qquad T=\frac{\Delta_r'(r_+)}{4\pi\sqrt{F(r_+)}}\,,\qquad \Omega=\frac{a\Xi}{F(r_+)}L(r_+)\,.
\ee
The four charges and their corresponding conjugate chemical potentials are given by
\bea
P_1&=&\frac{\mu}{2\sqrt{\Xi}}s_1c_1\,,\qquad Q_2=\frac{\mu}{2\sqrt{\Xi}}s_2c_2\,,\qquad P_3=\frac{\mu}{2\sqrt{\Xi}}s_3c_3\,,\qquad Q_4=\frac{\mu}{2\sqrt{\Xi}}s_4c_4\,,\cr
\Psi_1&=&\frac{2\mu}{\sqrt{\Xi}r_{1+}}[c_1s_1-a\alpha_1\Omega]\,,\qquad \Phi_2=\frac{2\mu}{\sqrt{\Xi}r_{2+}}[c_2s_2-a\alpha_2\Omega]\,,\cr
\Psi_3&=&\frac{2\mu}{\sqrt{\Xi}r_{3+}}[c_3s_3-a\alpha_3\Omega]\,,\qquad \Phi_4=\frac{2\mu}{\sqrt{\Xi}r_{4+}}[c_4s_4-a\alpha_4\Omega]\,.
\eea
Here, $r_{i+}=r_++2\mu s_i^2$ and $(P_1,\Psi_1)$ correspond to $A_{\1,1}$, $(Q_2,\Phi_2)$ to $A_{\1,2}$, $(P_3, \Psi_3)$ to $\mathcal{A}_{\1}^1$, and $(Q_4, \Phi_4)$ to $\mathcal{A}_{\1}^2$. The black hole mass and angular momentum are given, respectively, by
\be
M=\frac{\mu}{\Xi}\Big(1+\frac{1}{2}\sum_{i=1}^4s_i^2\Big)\,,\qquad J=\frac{\mu a}{\Xi}\big(c_{1234}-s_{1234}\big)\,.
\ee
For asymptotically AdS black holes, the thermodynamic pressure and its conjugate volume can be introduced within the framework of extended black hole thermodynamics \cite{Kastor:2009wy, Cvetic:2010jb}, and are given by
\bea
\mathcal{P}=\frac{3g^2}{8\pi}\,,\qquad V=\frac{\pi}{3\Xi}\Bigg[
P'(r_+)+a^2K'(r_+)+4\mu a^2\Big(1+\frac{1}{2}\sum_{i=1}^4s_i^2\Big)
\Bigg]\,.
\eea
These thermodynamic quantities can be shown to satisfy the first law
\be
dM=TdS+\Omega dJ+\Psi_1dP_1+\Phi_2dQ_2+\Psi_3dP_3+\Phi_4dQ_4+Vd\mathcal{P}.
\ee
The verification can be simplified by solving $\Delta_r(r_+)=0$ for $g^2$ and substitute the resulting expression into the above equation.

\subsection{BPS limit}

We now turn to the BPS limit. We focus on the branch with positive charge and positive angular momentum, which requires $a>0$ and $\delta_i>0$. Supersymmetry imposes a linear relation between the mass and conserved charges, as well as vanishing temperature:
\bea
T_*=0\,,\qquad M_*=P_{1,*}+Q_{2,*}+P_{3,*}+Q_{4,*}+gJ_*\,.\label{BPS condition}
\eea
Note that we use an asterisk subscript to denote quantities evaluated in the BPS limit. For convenience, we introduce two auxiliary quantities, $d$ and $\mathcal{D}$, defined by
\be
d=\sum_{i=1}^4\delta_i\,,\qquad \mathcal{D}=\prod_{1\leq i< j\leq4}\sinh(\delta_i+\delta_j)\,.
\ee
In the BPS limit, the mass and rotation parameters $(\mu, a)$, and the horizon radius $r_+$ are all determined by the four charge parameters $\delta_i$:
\bea
\mu_*=\frac{\sinh d}{2g\sqrt{\mathcal{D}\cosh d}}\,,\qquad a_*=\frac{1}{g\cosh d}\,,\qquad r_*=\frac{2\mu_*}{\sinh d}(c_1s_{\bar{1}}+c_2s_{\bar{2}}+c_3s_{\bar{3}}+c_4s_{\bar{4}})\,.\label{BPS condition 2}
\eea
In this limit, the remaining thermodynamic quantities reduce to
\bea
P_{1,*}&=&\frac{1}{4g}\sqrt{\frac{\cosh d}{\mathcal{D}}}s_1c_1\,,\qquad Q_{2,*}=\frac{1}{4g}\sqrt{\frac{\cosh d}{\mathcal{D}}}s_2c_2\,,\qquad P_{3,*}=\frac{1}{4g}\sqrt{\frac{\cosh d}{\mathcal{D}}}s_3c_3\,,\cr
Q_{4,*}&=&\frac{1}{4g}\sqrt{\frac{\cosh d}{\mathcal{D}}}s_4c_4\,,\qquad J_*=\frac{1}{2g^2\sinh d}\sqrt{\frac{\cosh d}{\mathcal{D}}}(c_{1234}-s_{1234})\,,\cr
\mathcal{P}_* &=& \frac{3g^2}{8\pi}\,,\qquad V_*=\frac{4\pi J_*}{3g}\,,\qquad
\Omega_* = g\,,\qquad \Psi_{1,*}=\Phi_{2,*}=\Psi_{3,*}=\Phi_{4,*}=1\,.
\eea
Thus, the BPS solutions are fully characterized by the four independent charges, with the mass, angular momentum and entropy determined by them. To express these relations compactly, we introduce the following charge-related parameters:
\be
\{\mathcal{Q}_{1*},\mathcal{Q}_{2*},\mathcal{Q}_{3*},\mathcal{Q}_{4*}\}=\{P_{1*},Q_{2*},P_{3*},Q_{4*}\}\,,\qquad \sigma_n=\sum_{i_1<\cdots <i_n}\mathcal{Q}_{i_1*}\cdots \mathcal{Q}_{i_n*}\,.
\ee
We find that the angular momentum satisfies a quadratic equation
\be
\Big(\sigma_3-\frac{J_*}{16g}\Big)^2-\sigma_1\Big(\sigma_2+\frac{1}{16g^2}\Big)
\Big(\sigma_3-\frac{J_*}{16g}\Big)+\sigma_1^2\Big(\sigma_4+\frac{J_*^2}{64}\Big)=0\,.
\ee
The entropy is simply given by
\be
S_* = \frac{4\pi}{g} \sqrt{\frac{\sigma_3}{\sigma_1} - \frac{J_*}{16g\,\sigma_1}}\,.
\ee
These results obtained from the BPS limit of the general non-extremal solution are in full agreement with those reported in the literature \cite{Hristov:2018spe,Choi:2018fdc,Hristov:2019mqp, Cassani:2019mms}.

\subsection{CTCs}

Further constraints are required for physically acceptable rotating black holes. Asymptically AdS geometry in spheroidal geometry ensures a finite ergoregion. We examine whether the solution is free of naked closed timelike curves (CTCs), requiring $g_{\phi\phi}>0$ on and outside the horizon.  From \eqref{metric}, this condition reduces to $N>0$. We note that
\be
\partial_r N=\sum_{i<j<k}r_ir_jr_k+\frac{a^2}{2}(1+\cos^2\theta)\sum_ir_i+\mu a^2(2+\sum_is_i^2)\sin^2\theta >0\,.
\ee
It follows that
\be
N(r,\theta)\geq N(r_+,\theta)=\Xi^{-1}\Delta_\theta F(r_+)=\Xi^{-1}\Delta_\theta W^2|_{r=r_+,\theta=0}>0\,.
\ee
Therefore, we rigorously establish that, for \(r\ge r_+\), \(0<\theta<\pi\), \(g_{\phi\phi}>0\).

As shown in \cite{Cacciatori:2008ek,Hristov:2019mqp}, the BPS black hole admits an everywhere non-spacelike Killing vector $K_+=\partial_t+g\partial_\phi$, constructed from the Killing spinor. For the general non-extremal solution, we have
\be
K_+^2=-\frac{W \Delta_r\Delta_\theta}{\Xi N }
+ \frac{N \sin^2\theta}{\Xi W} \Big(\frac{a \Delta_\theta L}{N}-g\Big)^2\,.
\ee
This implies that \(K_+^2\leq 0\) and \(g_{\phi\phi}>0\) cannot simultaneously hold on the horizon unless the bracketed term vanishes. This is indeed ensured by the BPS condition $\Omega_*=g$,
which fixes the horizon angular velocity.

\section{Conclusion}

We constructed exact non-extremal rotating AdS$_4$ black hole solutions carrying four independent charges in gauged STU supergravity, more than two decades after the special pairwise-equal solution was obtained. We analyzed their global structure, derived the black hole thermodynamics, verified the first law, and obtain the BPS limit. The general non-extremal solution consists of six parameters, namely the mass, angular momentum and four independent charges. The BPS black hole is extremal and satisfies an additional constraint, leaving the four charges as the only remaining independent parameters. Its entropy agrees with the known microscopic result from the dual CFT, providing a nontrivial consistency check of the holographic description. We have also proved that the black hole solutions are free of CTCs on and outside the event horizon.

These solutions provide gravitational backgrounds for studying charged, rotating thermal states of the dual three-dimensional CFT beyond the supersymmetric sector. The four independent charges allow us to investigate the effects of unequal chemical potentials on thermodynamic stability and phase transitions in different ensembles. Future work could explore near-BPS thermodynamics and higher-derivative supergravity corrections to the black hole entropy, using the non-extremal solutions to examine the approach to the supersymmetric limit and thereby extend the tests of the AdS/CFT correspondence beyond the leading order.

\section*{Acknowledgement}

L.M.~is supported in part by National Natural Science Foundation of China (NSFC) grant No.~12447138, Postdoctoral Fellowship Program of CPSF Grant No.~GZC20241211, the China Postdoctoral Science Foundation under Grant No.~2024M762338, the National Key Research and Development Program No.~2022YFE0134300, and also by the Tianjin University Self-Innovation Fund New Inspiration Fund No.~2026XQM09934. H.L.~is supported in part by the NSFC grants No.~12375052 and No.~11935009. The work is also supported in part by the Tianjin University Self-Innovation Fund Extreme Basic Research Project Grant No.~2025XJ21-0007. We also made an extensive use of AI programs in both research and English proofreading.

\end{document}